\documentclass[12pt,preprint]{aastex}

\usepackage{subfigure}

\shorttitle{The full curvature effect}

\shortauthors{Qin}

\begin{document}

\title{The full curvature effect expected in early X-ray
afterglow emission of gamma-ray bursts}

\author{Y.-P. Qin\altaffilmark{1,2}}

\altaffiltext{1}{Center for Astrophysics, Guangzhou University,
Guangzhou 510006, P. R. China; ypqin@gzhu.edu.cn}

\altaffiltext{2}{Physics Department, Guangxi University, Nanning
530004, P. R. China}

\begin{abstract}
We explore the influence of the full curvature effect on the flux of
early X-ray afterglow of gamma-ray bursts (GRBs) in cases when the
spectrum of the intrinsic emission is a power-law. We find that the
well-known $t^{-(2+\beta)}$ curve is present only when the intrinsic
emission is extremely short or the emission arises from an
exponential cooling. The time scale of this curve is independent of
the Lorentz factor. The resulting light curve would contain two
phases when the intrinsic emission has a power-law spectrum and a
temporal power-law profile with infinite duration. The first phase
is a rapid decay one where the light curve well follows the
$t^{-(2+\beta)}$ curve. The second is a shallow decay phase where
the power-law index of the light curve is obviously smaller than
that in the first phase. The start of the shallow phase is strictly
constrained by the fireball radius, which in turn, can put a lower
limit to the latter. In the case when the temporal power-law
emission lasts a limited interval of time, there will be a third
phase after the $t^{-(2+\beta)}$ curve and the shallow decay phase,
which is much steeper than the shallow phase. As an example of
application, we fit the XRT data of GRB 050219A with our model and
show that the curvature effect alone can roughly account for this
burst. Although fitting parameters can not be uniquely determined
due to various choices of fitting, a lower limit of the fireball
radius of this burst can be estimated, which is $\sim 10^{14}cm$.
\end{abstract}


\section{Introduction}

The canonical X-ray afterglow light curve containing five components
after the prompt emission phase is a great finding of Swift
(Chincarini et al. 2005; Nousek et al. 2006; O'Brien et al. 2006;
Zhang et al. 2006; Zhang 2007). The first of the five is the
so-called ``steep decay phase'' which generally extends to $\sim
(10^{2}-10^{3})s$, with a temporal decay slope typically $-3$ or
much steeper (Vaughan et al. 2006; Cusumano et al. 2006; O'Brien et
al. 2006).

A hint in this phase suggesting the emission of high latitude
fireball surface is that it is typically smoothly connected to the
prompt emission phase (Tagliaferri et al. 2005; Barthelmy et al.
2005; Liang et al. 2006). Generally, the steep decay phase was
interpreted as a consequence of the so-called curvature effect
(Fenimore et al. 1996; Kumar \& Panaitescu 2000; Dermer 2004; Dyks
et al. 2005; Butler \& Kocevski 2007a; Liang et al. 2006; Panaitescu
et al. 2006; Zhang et al. 2006; Zhang et al. 2007). The curvature
effect is a combined effect that includes the delay of time and the
shifting of the intrinsic spectrum as well as other relevant factors
of an expanding fireball (see Qin et al. 2006 for a detailed
explanation). The effect was intensively studied recently in the
prompt gamma-ray phase, where the profile of the full light curve of
pulses, the spectral lags, the power-law relation between the pulse
width and energy, and the evolution of the hardness ratio curve are
concerned (Sari \& Piran 1997; Qin 2002; Ryde \& Petrosian 2002;
Kocevski et al. 2003; Qin \& Lu 2005; Shen et al. 2005; Lu et al.
2006; Peng et al. 2006; Qin et al. 2004, 2005, 2006).

As early as a decade ago, Fenimore et al. (1996) found that, due to
the curvature effect, light curves arising from the emission of an
infinitely thin shell would be a power-law of observational time
when the rest-frame photon number spectrum is a power-law and the
emission is within an infinitesimal time interval. In this case, the
two power-law indexes are
related by $\alpha=2+\beta$, where $\alpha$ is the light curve index and $%
\beta$ the spectral index. In concerning the X-ray afterglow
emission, Kumar \& Panaitescu (2000) also found that, due to the
curvature effect, the light curve of a shocked heated fireball shell
radiating with a power-law spectrum within the observational band
(i.e., the X-ray band in the early afterglow observation) is a
power-law of time as well and relation $\alpha=2+\beta$ holds in
this situation.

As revealed in Fig. 7 of Nousek et al. (2006), relation
$\alpha=2+\beta$ is roughly in agreement with the data in the steep
decay phase of some Swift bursts. However, the figure also shows
that real relations between the two indexes of some bursts
significantly deviate from the $\alpha=2+\beta$ curve. This might be
due to the ill re-setting of time that should be set to the real
time when the central engine restarts (see Liang et al. 2006). In
addition, more or less subtracting the underlying afterglow
contribution would lead to other values of the temporal index
$\alpha$ (for a detailed explanation, see Zhang 2007).

We notice that the derivation of relation $\alpha=2+\beta$ in
previous papers is based on the main part of the curvature effect.
Does it still hold (or, in what situation it would still hold) when
the full curvature effect is considered? This motivates our
investigation below. The structure of the paper is as follows. In
Section 2, we present a general analysis on the full curvature
effect in cases when the intrinsic emission is a power-law. In
Section 3, we discuss light curves of power-law emission associated
with several typical intrinsic temporal profiles. Presented in
Section 4 is an example of application of our model. Conclusions are
presented in the last section.

\section{Light curves of fireballs arising from the intrinsic emission with a power-law
spectrum}

Observation of the emission arising from an expending fireball would
be influenced by the delay of time of different areas of the
fireball surface, the variation of the intensity due to the growing
of the fireball radius, the variation of the time contracted factor
and the shifting of the intrinsic spectrum associated with the angle
to the line of sight. Taking all these factors into account, one
comes to a full knowledge of the so-called curvature effect (see
also Qin et al. 2006 for a detailed explanation). Consider a
constant expanding fireball shell emitting within proper time
interval $t_{0,\min }\leq t_{0}\leq t_{0,\max }$ and over the
fireball area confined by $\theta _{min}\leq \theta \leq \theta
_{max}$, where $\theta $ is the angle to the line sight. Assume that
the energy range of the emission is not limited. Following the same
approach adopted in Qin (2002) and Qin et al. (2004), one can verify
that the flux tensity expected by a distant observer measured at
laboratory time $t_{ob}$ is
\begin{equation}
f_{\nu }(t_{ob})=\frac{2\pi c^{2}\int_{\widetilde{t}_{0,\min }}^{%
\widetilde{t}_{0,\max }}I_{0,\nu }(t_{0},\nu
_{0})[(t_{0}-t_{0,c})\Gamma
+D/c-(t_{ob}-t_{c})][R_{c}/c+(t_{0}-t_{0,c})\Gamma (v/c)]^{2}dt_{0}}{%
D^{2}\Gamma ^{2}\{R_{c}/c-[D/c-(t_{ob}-t_{c})](v/c)\}^{2}},
\end{equation}%
where $\widetilde{t}_{0,\min }$ and $\widetilde{t}_{0,\max }$ are
determined by
\begin{equation}
\widetilde{t}_{0,\min }=\max \{t_{0,\min },\frac{%
t_{ob}-t_{c}-D/c+(R_{c}/c)\cos \theta _{\max }}{[1-(v/c)\cos \theta
_{\max }]\Gamma }+t_{0,c}\}
\end{equation}%
and
\begin{equation}
\widetilde{t}_{0,\max }=\min \{t_{0,\max },\frac{%
t_{ob}-t_{c}-D/c+(R_{c}/c)\cos \theta _{\min }}{[1-(v/c)\cos \theta
_{\min }]\Gamma }+t_{0,c}\},
\end{equation}%
respectively, and $\nu _{0}$ and $t_{0}$ are related by
\begin{equation}
\nu _{0}=\frac{R_{c}/c-[D/c-(t_{ob}-t_{c})](v/c)}{R_{c}/c+(t_{0}-t_{0,c})%
\Gamma (v/c)}\Gamma \nu .
\end{equation}%
The observation time is confined by
\begin{equation}
\begin{array}{l}
\lbrack 1-(v/c)\cos \theta _{\min }][(t_{0,\min }-t_{0,c})\Gamma
+t_{c}]+[t_{c}(v/c)-R_{c}/c]\cos \theta _{\min }+D/c\leq t_{ob} \\
\leq \lbrack 1-(v/c)\cos \theta _{\max }][(t_{0,\max
}-t_{0,c})\Gamma
+t_{c}]+[t_{c}(v/c)-R_{c}/c]\cos \theta _{\max }+D/c%
\end{array}%
.
\end{equation}%
Beyond this time interval, no photons of the emission are detectable
by the observer.

A power-law spectrum was commonly observed in early X-ray afterglow
especially in the steep decay phase (e.g., Vaughan et al. 2006;
Cusumano et al. 2006; O'Brien et al. 2006). In this paper we focus
our attention only on the case of the intrinsic emission with a
power-law spectrum which is expectable in the case of synchrotron
emission produced by shocks and was generally assumed in previous
investigations (e.g., Fenimore et al. 1996; Sari et al. 1998; Kumar
\& Panaitescu 2000). Let the intensity of the intrinsic emission be
$I_{0,\nu }(t_{0},\nu _{0})=I_{0}(t_{0})\nu _{0}^{-\beta }$ (Kumar
and Panaitescu 2000). One gets from equation (1) that
\begin{equation}
f_{\nu }(t_{ob})=\frac{2\pi c^{2}\nu ^{-\beta
}\int_{\widetilde{t}_{0,\min }}^{\widetilde{t}_{0,\max
}}I_{0}(t_{0})[R_{c}/c+(t_{0}-t_{0,c})\Gamma
v/c]^{2+\beta }[(t_{0}-t_{0,c})\Gamma +D/c-(t_{ob}-t_{c})]dt_{0}}{%
D^{2}(\Gamma v/c)^{2+\beta }(t_{ob}-t_{c}+R_{c}/v-D/c)^{2+\beta }},
\end{equation}%
where relation (4) is applied. Assigning
\begin{equation}
t\equiv t_{ob}-t_{c}+R_{c}/v-D/c,
\end{equation}%
one comes to
\begin{equation}
f_{\nu }(t)=\frac{2\pi c^{2}\nu ^{-\beta }}{D^{2}(\Gamma
v/c)^{2+\beta }t^{2+\beta }}\int_{\widetilde{t}_{0,\min
}}^{\widetilde{t}_{0,\max
}}I_{0}(t_{0})[R_{c}/c+(t_{0}-t_{0,c})\Gamma v/c]^{2+\beta
}[(t_{0}-t_{0,c})\Gamma +R_{c}/v-t]dt_{0},
\end{equation}%
with%
\begin{equation}
\widetilde{t}_{0,\min }=\max \{t_{0,\min
},\frac{t-R_{c}/v+(R_{c}/c)\cos \theta _{\max }}{[1-(v/c)\cos \theta
_{\max }]\Gamma }+t_{0,c}\},
\end{equation}%
\begin{equation}
\widetilde{t}_{0,\max }=\min \{t_{0,\max
},\frac{t-R_{c}/v+(R_{c}/c)\cos \theta _{\min }}{[1-(v/c)\cos \theta
_{\min }]\Gamma }+t_{0,c}\},
\end{equation}%
\begin{equation}
\nu _{0}=\frac{t}{R_{c}/v+(t_{0}-t_{0,c})\Gamma }\Gamma \nu ,
\end{equation}%
and
\begin{equation}
\begin{array}{l}
\lbrack 1-(v/c)\cos \theta _{\min }][(t_{0,\min }-t_{0,c})\Gamma
+t_{c}]+[t_{c}(v/c)-R_{c}/c]\cos \theta _{\min }+R_{c}/v-t_{c}\leq t \\
\leq \lbrack 1-(v/c)\cos \theta _{\max }][(t_{0,\max
}-t_{0,c})\Gamma
+t_{c}]+[t_{c}(v/c)-R_{c}/c]\cos \theta _{\max }+R_{c}/v-t_{c}%
\end{array}
.
\end{equation}
The meaning of $t$ defined by equation (7) can be revealed by
employing equation (8) in Qin et al. (2004) (where quantity $t$ is
now written as $t_{ob}$). According to quation (8) in Qin et al.
(2004), emission from $R_c = 0$ (this emission occurs at
$t_{\theta}=t_c$) corresponds to $t=0$; and emission from the area
of $\theta = 0$ from any $R_c$ (occurring at $t_{\theta}=t_c$) gives
rise to $t=(R_c/v)(1-\beta)\simeq(R_c/v)/2\Gamma^2$. Quantity
$(R_c/v)/2\Gamma^2$ is nothing but the traveling time of the
fireball surface from the explosion spot to $R_c$, contracted by
factor $1/2\Gamma^2$ since the area of $\theta=0$ moves towards the
observer with Lorentz factor $\Gamma$. Thus, $t=0$ is the moment
when photons emitted from $R_c=0$ reach the observer. Even for
$t_{\theta}=t_c$, one would have $t>0$ if $R_c>0$. The emission time
$t_{\theta}=t_c$ does not mean that photons are radiated at $R_c=0$.
In stead, it means that these photons are emitted from the surface
of the fireball with radius $R_c$ which is measured at $t_c$ (see
Qin 2002 Appendix A).

Note that when the power-law range is limited, then it would
constrain the integral limits $\widetilde{t}_{0,\min }$ and
$\widetilde{t}_{0,\max }$ which are different from equations (2) and
(3), or (9) and (10) (see Qin 2002). In the following, we adopt the
Kumar \& Panaitescu (2000)'s assumption: the intrinsic emission is a
strict power-law within the energy range corresponding to the
observed energy channel. Thus, equations (2) and (3), or (9) and
(10) are applicable.

According to (9) and (10), $\int_{\widetilde{t}_{0,\min }}^{\widetilde{t}%
_{0,\max }}I_{0}(t_{0})[R_{c}/c+(t_{0}-t_{0,c})\Gamma v/c]^{2+\beta
}[(t_{0}-t_{0,c})\Gamma +R_{c}/v-t]dt_{0}$ is only a function of $t$. Let%
\begin{equation}
h(t)\equiv \int_{\widetilde{t}_{0,\min }}^{\widetilde{t}_{0,\max
}}I_{0}(t_{0})[R_{c}/c+(t_{0}-t_{0,c})\Gamma v/c]^{2+\beta
}[(t_{0}-t_{0,c})\Gamma +R_{c}/v-t]dt_{0}.
\end{equation}%
Equation (8) could then be written as
\begin{equation}
f_{\nu }(t)=\frac{2\pi c^{2}}{D^{2}(\Gamma v/c)^{2+\beta }}%
h(t)t^{-(2+\beta )}\nu ^{-\beta }.
\end{equation}%
It shows that, in the case that the power-law intrinsic radiation
intensity $I_{0,\nu }(t_{0},\nu _{0})=I_{0}(t_{0})\nu _{0}^{-\beta
}$ holds within the energy range which corresponds to the observed
energy channel due to the Doppler shifting, a power-law spectrum
will also hold within the observed channel and the index will be
exactly the same as that in the intrinsic spectrum.

Taking factor $h(t)$ as a constant, equation (14) gives rise to
\begin{equation}
f_{\nu }(t)\propto t^{-(2+\beta )}\nu ^{-\beta }.
\end{equation}%
This is the well-known flux density associated with the curvature
effect, which reveals the relation between the temporal and spectral
power-law indexes: $\alpha =2+\beta $, where $\alpha$ is the
temporal index (e.g., when assuming $f_{\nu }(t)\propto t^{-\alpha
}$) (see Fenimore et al. 1996; Kumar \& Panaitescu 2000).

\section{Time factors other than the power-law function}

Let us consider an intrinsic emission with a $\delta $ function of
time. In this situation, effects arising from the duration of real
intrinsic emission will be omitted and therefore those merely coming
from the expanding motion of the fireball surface will be clearly
seen.

Not losing generality, we assume that $I_{0}(t_{0})=I_{0}\delta
(t_{0}-t_{0,c})$ and take $\theta _{\min }=0$ and $\theta _{\max
}=\pi /2$ (this corresponds to the half fireball surface facing us,
which will be taken throughout this paper). One then gets from (12)
that
\begin{equation}
R_{c}/v-R_{c}/c\leq t\leq R_{c}/v.
\end{equation}
Within the observation time confined by (16), the integral (13)
becomes
\begin{equation}
h(t)= I_{0}(R_{c}/c)^{2+\beta }(R_{c}/v-t).
\end{equation}%
Therefore,%
\begin{equation}
f_{\nu }(t)=\frac{2\pi c^{2}I_{0}(R_{c}/\Gamma v)^{2+\beta }}{D^{2}}%
(R_{c}/v-t)t^{-(2+\beta )}\nu ^{-\beta }.
\end{equation}%
When $t$ is beyond the time range confined by (16), $h(t)=0$ and
then $f_{\nu }(t)=0$.

Based on Qin (2002) and Qin et al. (2004), one can check that the
term $[(t_{0}-t_{0,c})\Gamma +D/c-(t_{ob}-t_{c})]$ in equation (1),
or the term $[(t_{0}-t_{0,c})\Gamma +R_{c}/v-t]$ in equation (13),
comes from the projected factor of the infinitesimal fireball
surface area in the angle concerned (say, $\theta$) to the distant
observer, which is known as $cos \theta$. This term becomes
$R_{c}/v-t$ for a $\delta$-function temporal radiation when adopting
the new time definition (7). Corresponding to larger observation
times, the line of sight angles of the emitted areas are larger, and
then the term $cos \theta$ becomes smaller.

Note that when one considers only a very small cone towards the
observer, this term could be ignored since it varies very mildly
within the angle range close to $\theta = 0$. However, what we
discuss here is the steep decay phase of early X-ray afterglow which
was generally assumed to arise from high latitude emission. In this
situation, the variation of this term would be significant.

According to (11), another noticeable term,
$[R_{c}/c+(t_{0}-t_{0,c})\Gamma v/c]$, in equation (1) or (13)
reflects the shifting of frequency. Equation (11) suggests that the
flux observed at frequency $\nu$ and time $t$ will be contributed by
rest-frame photons of frequency $\nu_0$ emitted at proper time $t_0$
(note that the flux will also be contributed by rest-frame
photons of other frequency $\nu_{0}^{\prime}$ emitted at other proper time $%
t_{0}^{\prime}$ so long as they satisfy equation (11), and the value
of the
flux is determined by all these possible photons). Quantity $%
[R_{c}/c+(t_{0}-t_{0,c})\Gamma v/c]$ is a shifting factor of the
frequency when observation time $t$ is fixed. This term is
independent of observation
time, but due to its coupling with $(t_{0}-t_{0,c})\Gamma$ in the term of $%
[(t_{0}-t_{0,c})\Gamma +D/c-(t_{ob}-t_{c})]$, it might also affect
the light curve.

Similarly, the term $I_{0}(t_{0})$ might also play a role due to its
coupling with $(t_{0}-t_{0,c})\Gamma$ in the term
$[(t_{0}-t_{0,c})\Gamma +D/c-(t_{ob}-t_{c})]$.

The last factor affecting the profile of light curves is the
integral range of equation (13). The integral range might differ
from time to time since the fireball surface area that sends photons
to the observer, which are observed at time $t$, might
change with time. Shown in (9) and (10), both $\widetilde{t}_{0,\min }$ and $%
\widetilde{t}_{0,\max }$ are determined by observation time $t$. A
time dependent integral range in equation (13) probably could make
the decay phase of the light curve deviate from a strict power-law
one.

\begin{figure}[tbp]
\begin{center}
\includegraphics[width=5in,angle=0]{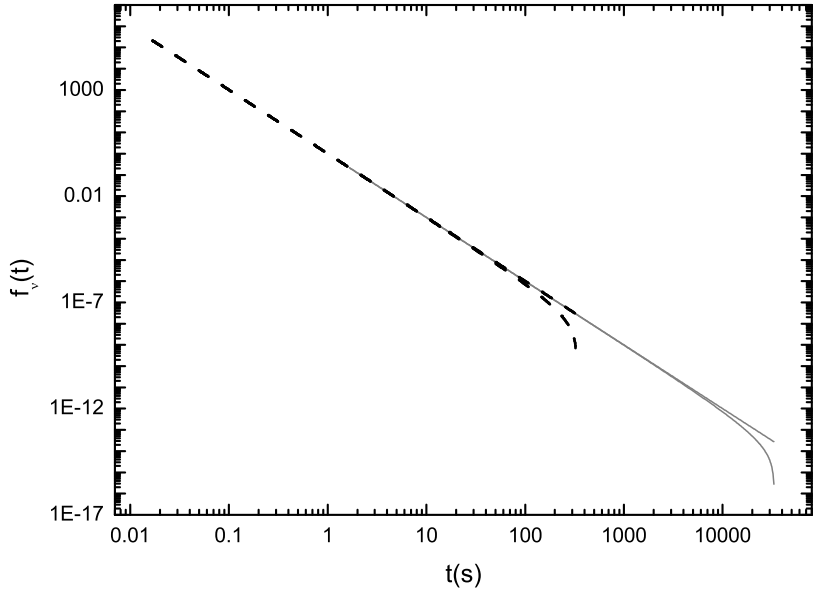}
\end{center}
\caption{Light curves $I_{\nu,\delta}[1-t/(R_{c}/v)]t^{-(2+\beta )}$
(lower lines) and $I_{\nu,\delta}t^{-(2+\beta )}$ (upper lines)
associated with $R_{c}\simeq 10^{15}cm$ (solid lines) and
$R_{c}\simeq 10^{13}cm$ (dashed lines). Note that the lower and
upper lines are overlapped in the main domain of the corresponding
light curves.} \label{Fig. 1}
\end{figure}

In the following, we show how these time factors affect the decay
phase of light curves which arise from the intrinsic emission with a
power-law spectrum.

\subsection{In the case of the temporal profile of the intrinsic emission being a $\delta$-function of time}

We first consider the case of the temporal profile of the intrinsic
emission being a $\delta$-function of time. The light curve arising
from this emission is $I_{\nu,\delta}[1-t/(R_{c}/v)]t^{-(2+\beta )}$
according to (18). We take $I_{\nu,\delta}=1$ and $\beta=1$ to plot
the curves. We consider the fireball radius with
$R_{c}/v=(1/3)10^{5}s$ and $ R_{c}/v=(1/3)10^{3}s $ which correspond
to two typical radius $R_{c}\simeq 10^{15}cm$ and $R_{c}\simeq
10^{13}cm$ respectively (see Ryde \& Petrosian 2002).

Shown in Fig. 1 are the light curves of
$I_{\nu,\delta}[1-t/(R_{c}/v)]t^{-(2+\beta )}$ and
$I_{\nu,\delta}t^{-(2+\beta )}$ associated with $R_{c}\simeq
10^{15}cm$ and $R_{c}\simeq 10^{13}cm$. We find that, in the case of
very short intrinsic emission, although the $R_{c}/v-t$ term in
equation (18) plays a role in the decay phase, the temporal curve
well follow a power-law in the main domain of the phase. Following
the power-law curve is a tail falling off speedily due to the effect
of the $R_{c}/v-t$ term. A remarkable feature revealed by the figure
is that the power-law decay time is solely determined by and very
sensitive to the radius of the fireball and the power-law range
itself can tell how large is a fireball radius. For example, a
power-law range being found to extend to $100s$ must be larger than
$10^{13}cm$ and that being found to extend to $10000s$ must be
larger than $10^{15}cm$. The conclusion is surprisingly to be
independent of the Lorentz factor. Note that this conclusion holds
when the intrinsic emission is extremely short so that its temporal
profile can be treated as a $\delta$-function. As illustrated in
Fig. 1, a strict $t^{-(2+\beta )}$ curve followed by a speedily
falling off tail is a feature of extremely short intrinsic emission.
When this feature is observed, one can estimate the fireball radius
merely from the time scale of the power-law decay phase so long as
the spectrum is a power-law and the relation of $\alpha =2+\beta $
holds.

\subsection{In the case of the temporal profile of the intrinsic emission being an exponential
function of time}

Second, let us consider the intrinsic emission with its temporal
profile being an exponential of time and check if the resulting
light curve is different from that arising from the
$\delta$-function emission. We ignore the contribution from the rise
phase of the emission of shocks (it corresponds to the situation
when the rise time is extremely short). The intrinsic decaying light
curve with an exponential form is assumed to be:
$I_{0}(t_{0})=I_{0}exp[-(t_{0}-t_{0,c})/ \sigma _{d}]$ for
$t_{0}>t_{0,c}$. Equation (14) now becomes
\begin{equation}
f_{\nu }(t)=I_{e}h_{e}(t)t^{-(2+\beta )}\nu ^{-\beta },
\end{equation}%
with
\begin{equation}
h_{e}(t)=\int_{\widetilde{t}_{0,\min }}^{\widetilde{t}_{0,\max }}\frac{%
[1+(t_{0}-t_{0,c})\Gamma v/R_{c}]^{2+\beta }[(t_{0}-t_{0,c})\Gamma
v/R_{c}+1-tv/R_{c}]dt_{0}}{exp[(t_{0}-t_{0,c})/\sigma _{d}]}\qquad
\qquad (t_{0}>t_{0,c}),
\end{equation}%
\begin{equation}
\widetilde{t}_{0,\min }=\max \{t_{0,c},\frac{t-R_{c}/v}{\Gamma }%
+t_{0,c}\}
\end{equation}%
and%
\begin{equation}
\widetilde{t}_{0,\max }=\frac{t-R_{c}/v+R_{c}/c}{(1-v/c)\Gamma }%
+t_{0,c},
\end{equation}%
where $I_{e}$ is a constant and observation time $t$ is confined by
\begin{equation}
R_{c}/v-R_{c}/c\leq t.
\end{equation}

Not losing generality, we take $t_{0,c}=0$. Equations (20)-(22) then
become
\begin{equation}
h_{e}(t)=\int_{\widetilde{t}_{0,\min }}^{\widetilde{t}_{0,\max }}\frac{%
(1+t_{0}\Gamma v/R_{c})^{2+\beta }(t_{0}\Gamma v/R_{c}+1-tv/R_{c})dt_{0}}{%
exp(t_{0}/\sigma _{d})}\qquad \qquad (t_{0}>0),
\end{equation}%
\begin{equation}
\widetilde{t}_{0,\min }=\max \{0,\frac{t-R_{c}/v}{\Gamma }\}
\end{equation}%
and%
\begin{equation}
\widetilde{t}_{0,\max }=\frac{t-R_{c}/v+R_{c}/c}{(1-v/c)\Gamma }.
\end{equation}

Here, we take $I_{e}=1$, $\beta =1$, and adopt
$R_{c}/v=(1/3)10^{5}s$ to plot the light curves. For the Lorentz
factor, we take $\Gamma =10$ and $\Gamma =100$, respectively.

\begin{figure}[tbp]
\begin{center}
\includegraphics[width=5in,angle=0]{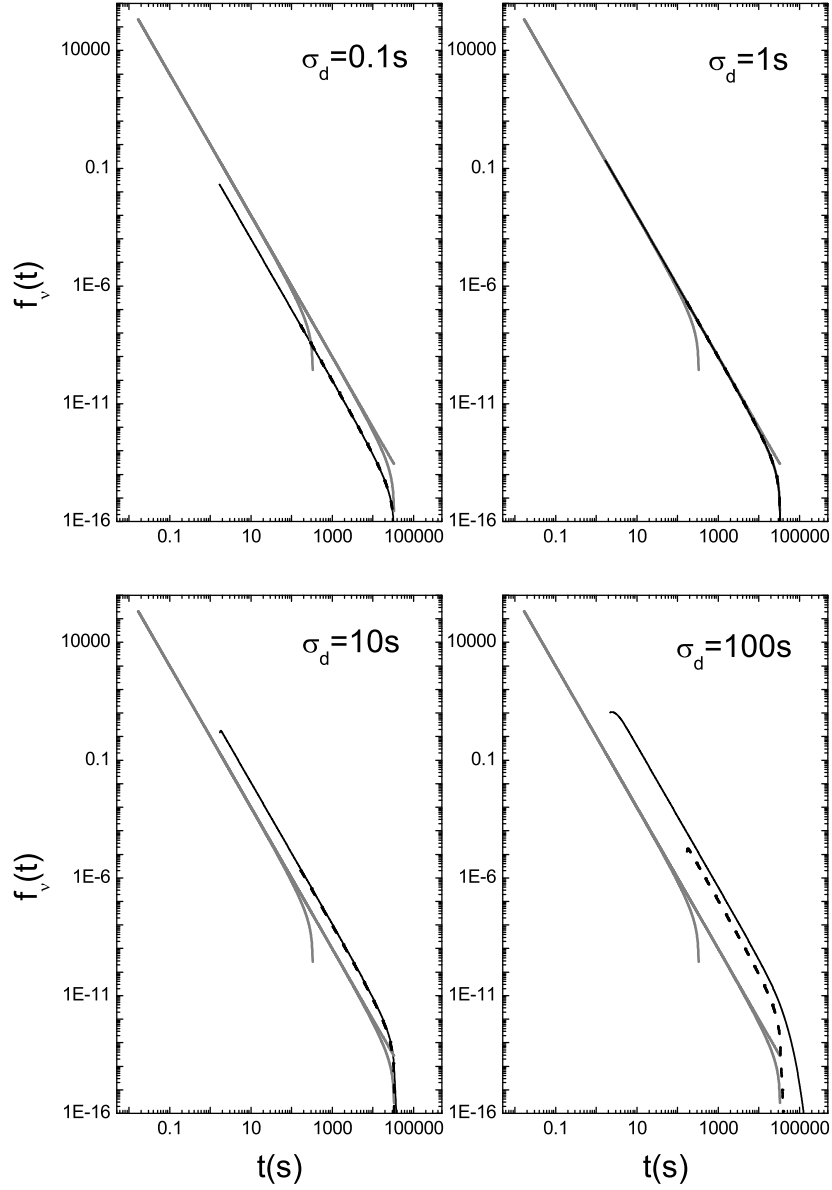}
\end{center}
\caption{Light curves (solid lines for $\Gamma=100$; dashed lines
for $\Gamma=10$) arising from the intrinsic emission with its
temporal profile being an exponential function of time (see equation
(19)), plotted in cases of $\sigma_{d}=0.1s$, $1s$, $10s$ and $100s$
respectively, where $h_{e}(t)$ is determined by (24). For the sake
of comparison, those lines in Fig. 1 are also plotted (the grey
color lines).} \label{Fig. 2}
\end{figure}

Shown in Fig. 2 are the light curves plotted with different values
of the width of the exponential function ($\sigma_{d}=0.1s$, $1s$,
$10s$ and $100s$). The light curves are quite similar to those
arising from the intrinsic emission with its temporal profile being
a $\delta$-function of time (see Fig. 1, where a feature of a
$t^{-(2+\beta )}$ curve followed by a speedily falling off tail is
observed). Due to the contribution of the exponential decay curve of
the intrinsic emission, the range of light curves is sightly larger
than that in the case of the intrinsic emission with a
$\delta$-function of time (this can be observed when the width is
large enough; see the lower right panel of Fig. 2). This is
understandable since after the width the emission of an exponential
function dies away rapidly and therefore its contribution can be
ignored.

In the case when both the width of the exponential function emission
and the Lorentz factor of the fireball are large, the resulting
light curve would obviously deviate from that arising from the
$\delta$-function emission in the domain of the falling off tail,
where the slope of the tail of the former light curve becomes
obviously mild (see the lower right panel of Fig. 2). Besides this,
no other characteristics can distinguish the tow kinds of light
curve.

\subsection{In the case of the temporal profile of the intrinsic emission being a power-law function of
time}

Third, we check if an observed light curve arising from the emission
with a power-law spectrum has something to do with the intrinsic
decaying behavior when the decay curve is a power-law of time. Here,
we also ignore the contribution from the rise phase of the emission
of shocks, and then consider only an intrinsic power-law decay
emission (this will occur when the cooling is a power law).

\subsubsection{When the power-law decay time is infinity}

Assuming that the power-law decay time is infinity, the intrinsic
decaying light curve is taken as
$I_{0}(t_{0})=[(t_{0}-t_{0,c})/(t_{0,0}-t_{0,c})]^{-\alpha _{0}}$
for $t_{0}>t_{0,0}$, where $t_{0,0}>t_{0,c}$ is a constant which is
the time when the power-law decay emission begins. In this case,
equation (14) becomes
\begin{equation}
f_{\nu }(t)=I_{p}h_{p}(t)t^{-(2+\beta )}\nu ^{-\beta },
\end{equation}%
with
\begin{equation}
h_{p}(t)=\int_{\widetilde{t}_{0,\min }}^{\widetilde{t}_{0,\max }}\frac{%
[1+(t_{0}-t_{0,c})\Gamma v/R_{c}]^{2+\beta }[(t_{0}-t_{0,c})\Gamma
v/R_{c}+1-tv/R_{c}]}{[(t_{0}-t_{0,c})/(t_{0,0}-t_{0,c})]^{\alpha _{0}}}%
dt_{0}\qquad (t_{0}>t_{0,0}),
\end{equation}%
\begin{equation}
\widetilde{t}_{0,\min }=\max \{t_{0,0},\frac{t-R_{c}/v}{\Gamma
}+t_{0,c}\}
\end{equation}%
and%
\begin{equation}
\widetilde{t}_{0,\max }=\frac{t-R_{c}/v+R_{c}/c}{(1-v/c)\Gamma
}+t_{0,c},
\end{equation}
where $I_{p}$ is a constant and observation time $t$ is confined by
\begin{equation}
(t_{0,0}-t_{0,c})(1-v/c)\Gamma -R_{c}/c+R_{c}/v\leq t.
\end{equation}%
Not losing generality, we take $t_{0,c}=0$. Equations (28)-(31) then
become
\begin{equation}
h_{p}(t)=\int_{\widetilde{t}_{0,\min }}^{\widetilde{t}_{0,\max
}}(1+t_{0}\Gamma v/R_{c})^{2+\beta }(1+t_{0}\Gamma
v/R_{c}-tv/R_{c})(t_{0}/t_{0,0})^{-\alpha _{0}}dt_{0}\qquad \qquad
(t_{0}>t_{0,0}),
\end{equation}%
\begin{equation}
\widetilde{t}_{0,\min }=\max \{t_{0,0},\frac{t-R_{c}/v}{\Gamma }\},
\end{equation}%
\begin{equation}
\widetilde{t}_{0,\max }=\frac{t-R_{c}/v+R_{c}/c}{(1-v/c)\Gamma }
\end{equation}%
and%
\begin{equation}
(1-v/c)\Gamma t_{0,0}-R_{c}/c+R_{c}/v\leq t.
\end{equation}

Here, we take $I_{p}\nu^{-\beta}=1$ and $\beta =1$ to plot the light
curves. For the Lorentz factor, we take $\Gamma=10$ and
$\Gamma=100$. We consider two typical values of the fireball radius
$ R_{c} = 10^{15}cm$ and $R_{c} = 10^{13}cm$. For the intrinsic
temporal power-law index, we take $\alpha _{0}=2$, $2.5$, $3$, $4$
and $5$ respectively, and for the time when the power-law decay
emission begins we take $t_{0,0}=0.01s$, $0.1s$ and $1s$
respectively.

\begin{figure}
  \centering
  \subfigure[]{
    \label{fig:subfig:a} 
    \includegraphics[width=2.2in]{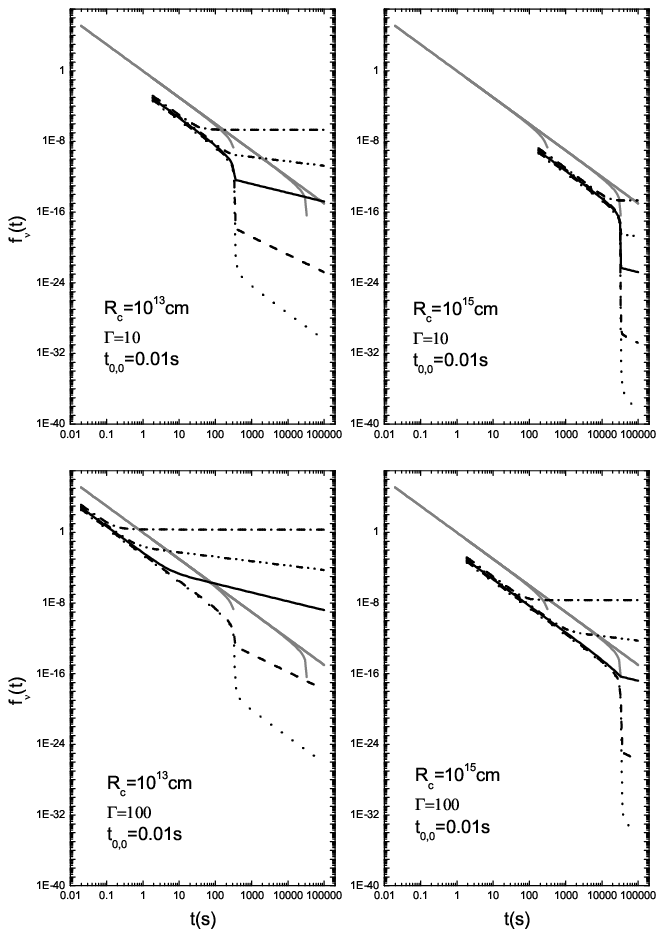}}
  \subfigure[]{
    \label{fig:subfig:b} 
    \includegraphics[width=2.2in]{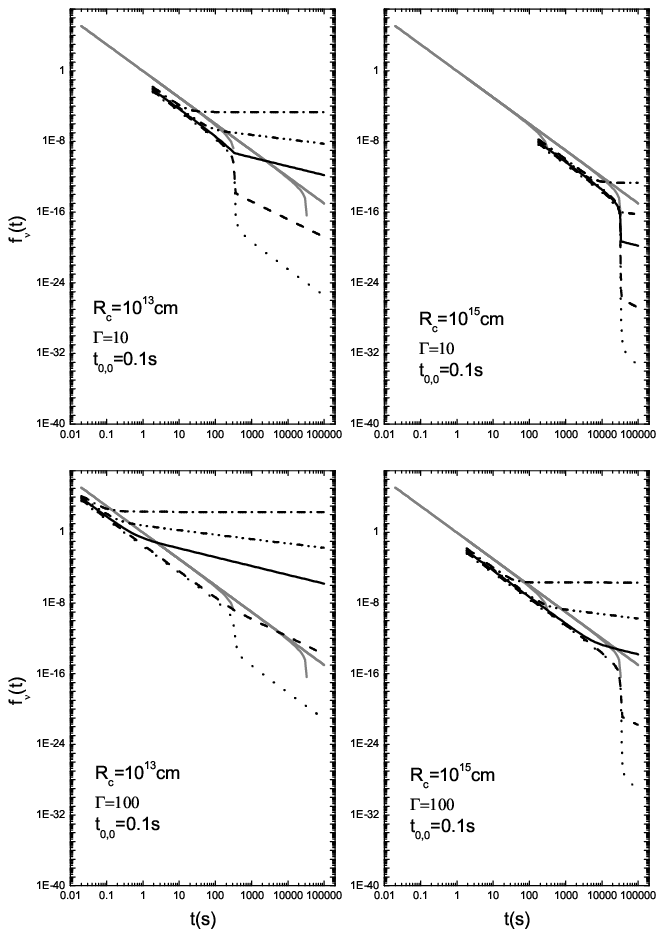}}
  \subfigure[]{
    \label{fig:subfig:c} 
    \includegraphics[width=2.2in]{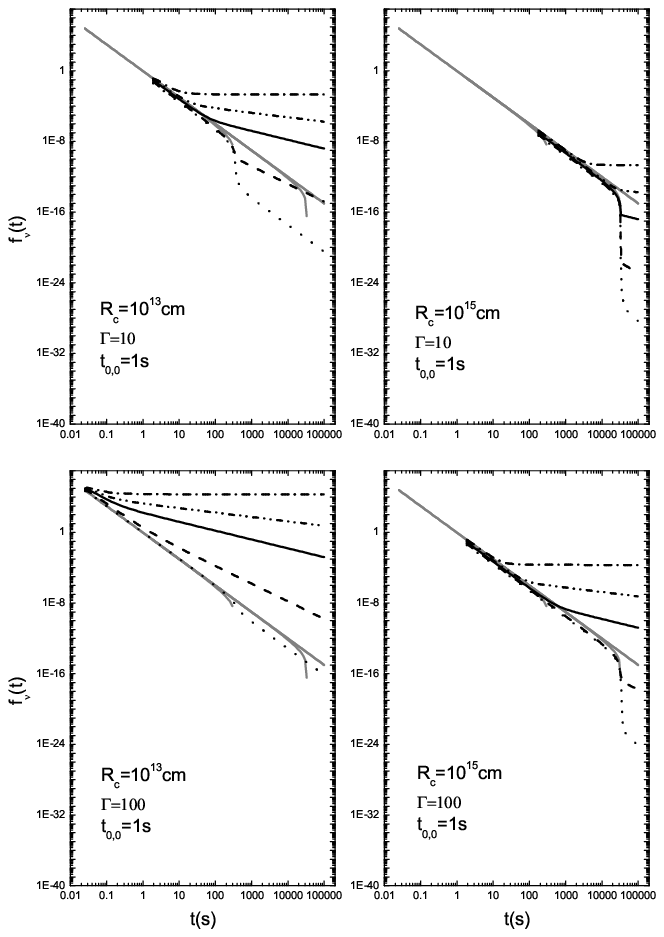}}
 \caption{Light curves arising from the intrinsic emission with its
temporal profile being a power-law function of time and the
power-law decay time being infinity (see equation (27)), plotted in
cases of $t_{0,0}=0.01s$ (sub-figure a), $0.1s$ (sub-figure b) and
$1s$ (sub-figure c) respectively, where $h_{p}(t)$ is determined by
(32). The upper and lower panels of each sub-figure correspond to
$\Gamma = 10$ and $\Gamma = 100$ respectively and the left and right
panels of each sub-figure correspond to $R_{c} = 10^{13}cm$ and
$R_{c} = 10^{15}cm$ respectively. Five kinds of black line stand for
five different intrinsic temporal power-law indexes. They are
(counting the five black lines for each panel from the top to the
bottom): the dash dot line for $\alpha _{0}=2$ ($\alpha
_{0}=1+\beta$), the dash dot dot line for $\alpha _{0}=2.5$ ($\alpha
_{0}=1.5+\beta$), the solid line for $\alpha _{0}=3$ ($\alpha
_{0}=2+\beta$), the dash line for $\alpha _{0}=4$ ($\alpha
_{0}=3+\beta$) and the dot line for $\alpha _{0}=5$ ($\alpha
_{0}=4+\beta$). For the sake of comparison, those lines in Fig. 1
are also plotted (the grey color lines).}
\end{figure}

The corresponding light curves are displayed in Fig. 3. Due to the
contribution of $h_{p}(t)$, some new features are observed. There
exist two kinds of light curve: a) a $t^{-(2+\beta)}$ curve followed
by a shallow decay curve with its index being obviously smaller than
$2+\beta$ (type I); b) a $t^{-(2+\beta)}$ curve followed by a very
steep decay phase (shown as a ``cutoff'' curve) and then a shallow
decay curve with its index being smaller than $2+\beta$ (type II).
The curve of type II tends to appear in cases when the intrinsic
temporal power-law index is large, the Lorentz factor is small and
the onset of the intrinsic temporal power-law is early (comparing
left panels of sub-figures a, b and c; or comparing right panels of
the sub-figures). The very steep decay curve appears very close to
the time position marked by that in the light curve arising from the
$\delta$ function emission (see the gray color lines in the figure)
(in fact, relative to the latter, the former shifts to slightly
larger time scales). This means that the time position of the very
steep decay phase of the light curve of type II is mainly determined
by the radius of the fireball, which can serve as an indicator of
the latter (see also the discussion in the two previous
subsections). For the light curve of type I, the start of the
shallow decay phase can appear from very early time scale to around
300s for the fireball with radius $R_{c} = 10^{13}cm$, depending on
the intrinsic temporal power-law index $\alpha_0$, the Lorentz
factor $\Gamma$ and the onset time $t_{0,0}$ of the intrinsic
temporal power-law (see the left panels of the sub-figures a, b and
c). The smaller values of $\alpha_0$, $\Gamma$ and $t_{0,0}$, the
larger time scale of the start of the shallow decay phase. For the
fireball with radius $R_{c} = 10^{15}cm$, conclusions drawn from
type I light curves remain the same, except that the maximum of the
start time of the shallow decay phase can appear at around 3000s. In
both types I and II, the slope of the shallow decay curve increases
with the increasing of $\alpha_0$.

Revealed in the left lower panel of Fig. 3c, as a special case of
type I, some light curves appear to be a single power-law one with
their indexes significantly smaller than $2+\beta$. They are in fact
the shallow decay phase of the corresponding light curves. The onset
of the phase shifts to much smaller time scales due to the larger
values of the Lorentz factor $\Gamma$ and the onset time $t_{0,0}$
of the intrinsic temporal power-law for a given value of the
fireball radius.

\subsubsection{When the power-law decay time is limited}

One might notice that there is no upper limit of the intrinsic
power-law decay emission considered above. Let us put an upper limit
to the intrinsic emission and then check if it could give rise to
other noticeable features on the observed light curves. The
intrinsic decaying light curve is assumed
to be $I_{0}(t_{0})=[(t_{0}-t_{0,c})/(t_{0,0}-t_{0,c})]^{-\alpha _{0}}$ for $%
t_{0,0}<t_{0}<t_{0,\max }$. Also, we take $t_{0,c}=0$. In this
situation, equations (27), (32) and (33) hold, while equations (34)
and (35) are
replaced by%
\begin{equation}
\widetilde{t}_{0,\max }=\min \{t_{0,\max },\frac{t-R_{c}/v+R_{c}/c}{%
(1-v/c)\Gamma }\},
\end{equation}
\begin{equation}
(1-v/c)\Gamma t_{0,0}-R_{c}/c+R_{c}/v\leq t\leq t_{0,\max }\Gamma
+R_{c}/v.
\end{equation}

Parameters adopted in producing Fig. 3 are also adopted here to
create the light curves. Among those studied in Fig. 3, we consider
only the following four cases: $\Gamma =10$ and $t_{0,0}=0.01s$ (see
the upper panels of Fig. 3a); $\Gamma =10$ and $t_{0,0}=0.1s$ (see
the upper panels of Fig. 3b); $\Gamma =100$ and $t_{0,0}=0.1s$ (see
the lower panels of Fig. 3b); $\Gamma =100$ and $t_{0,0}=1s$ (see
the lower panels of Fig. 3c). For the new parameter, we take
$t_{0,\max }= t_{0,0}+0.01R_{c}/c$, $t_{0,\max }= t_{0,0}+R_{c}/c$
and $t_{0,\max }= t_{0,0}+100R_{c}/c$ respectively. The
corresponding light curves are displayed in Figs. 4-7 which
correspond to the four cases respectively.

\begin{figure}[tbp]
\begin{center}
\includegraphics[width=5in,angle=0]{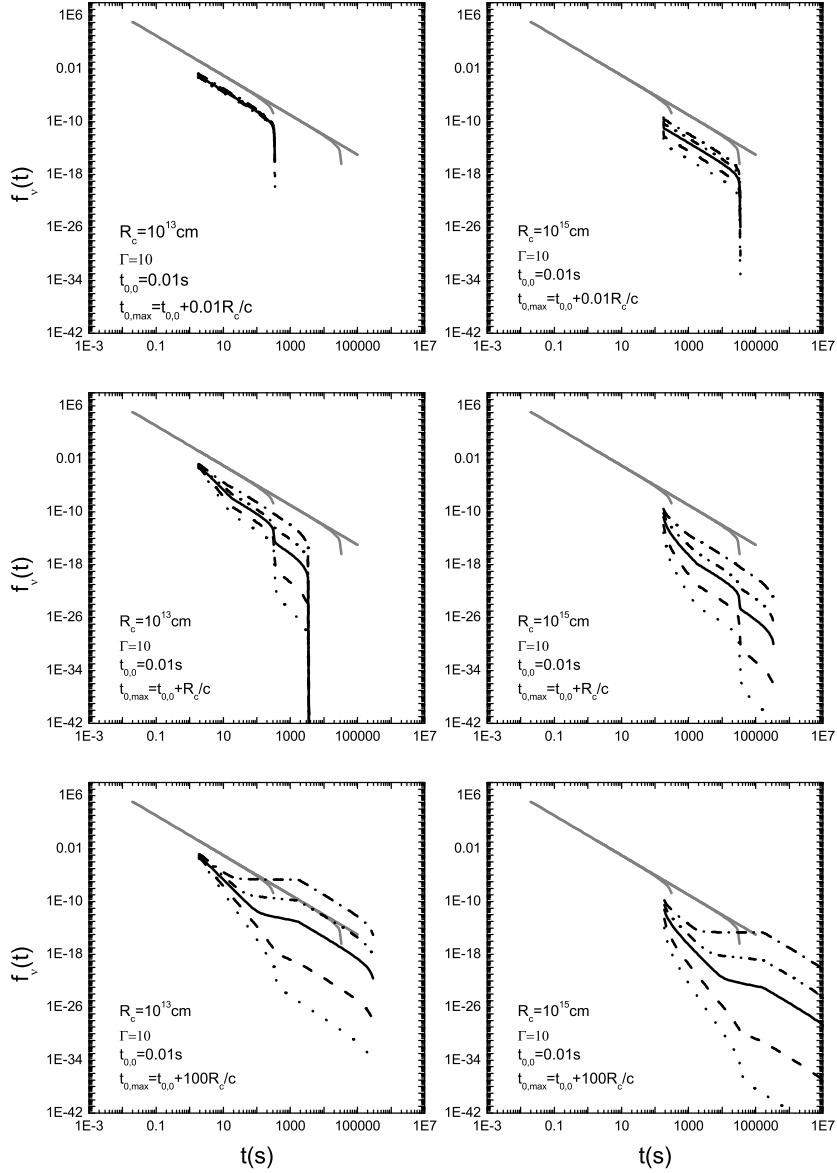}
\end{center}
\caption{Light curves arising from the intrinsic emission with its
temporal profile being a power-law function of time and with a
limited duration, plotted in the case of $\Gamma =10$ and
$t_{0,0}=0.01s$. Here, we consider two values of the fireball radius
and three time scales of the duration of the power-law emission (see
the description in each panel). The equations are the same as that
adopted in Fig. 3, except that we use equations (36) and (37) to
replace equations (34) and (35), respectively. The symbols are the
same as that in Fig. 3.} \label{Fig. 1}
\end{figure}

\begin{figure}[tbp]
\begin{center}
\includegraphics[width=5in,angle=0]{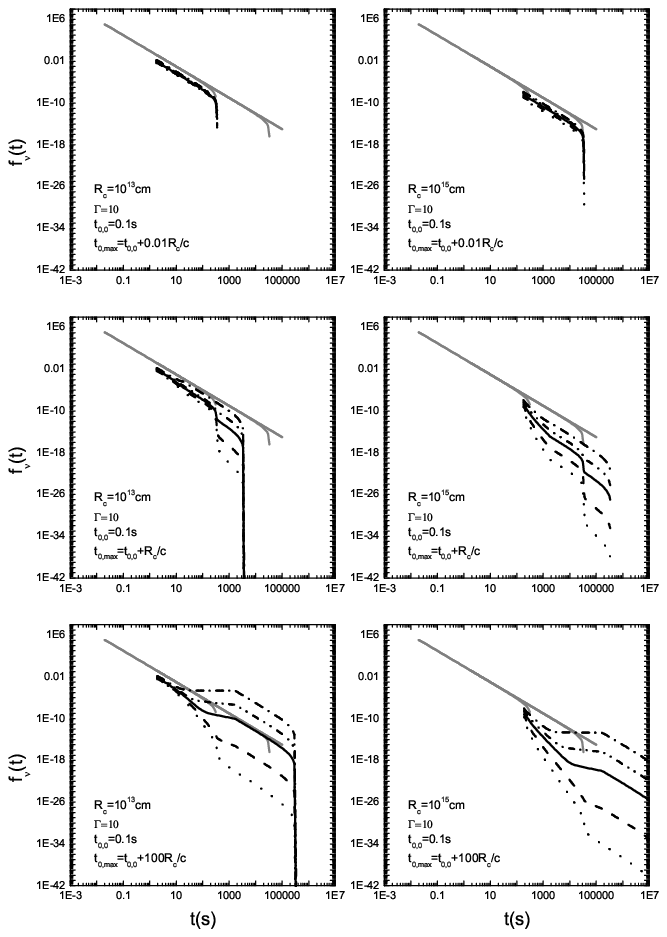}
\end{center}
\caption{Light curves of Fig. 4 replaced by those produced in the
case of $\Gamma =10$ and $t_{0,0}=0.1s$.} \label{Fig. 1}
\end{figure}

\begin{figure}[tbp]
\begin{center}
\includegraphics[width=5in,angle=0]{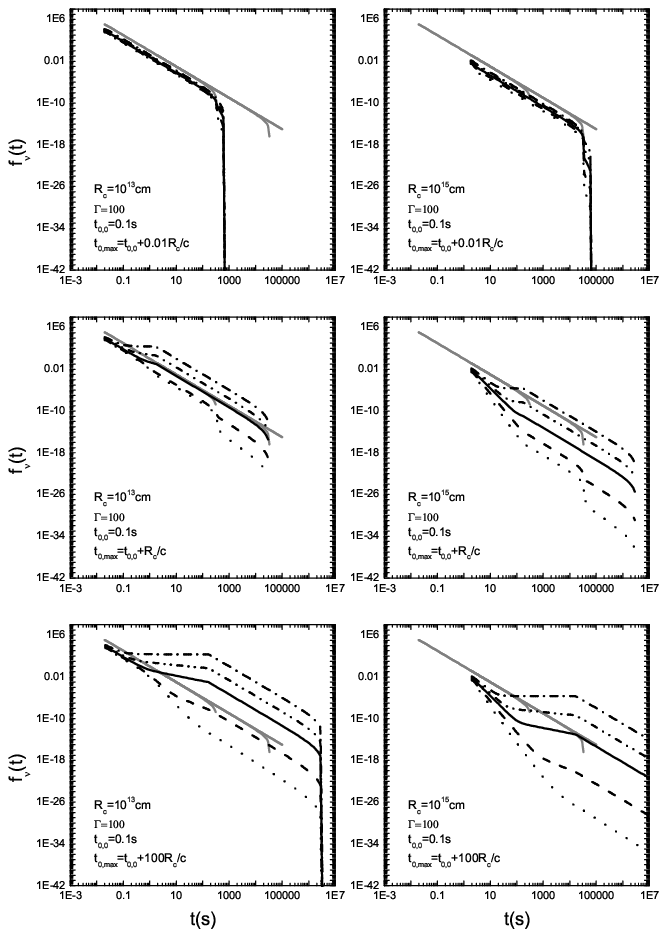}
\end{center}
\caption{Light curves of Fig. 4 replaced by those produced in the
case of $\Gamma =100$ and $t_{0,0}=0.1s$.} \label{Fig. 1}
\end{figure}

\begin{figure}[tbp]
\begin{center}
\includegraphics[width=5in,angle=0]{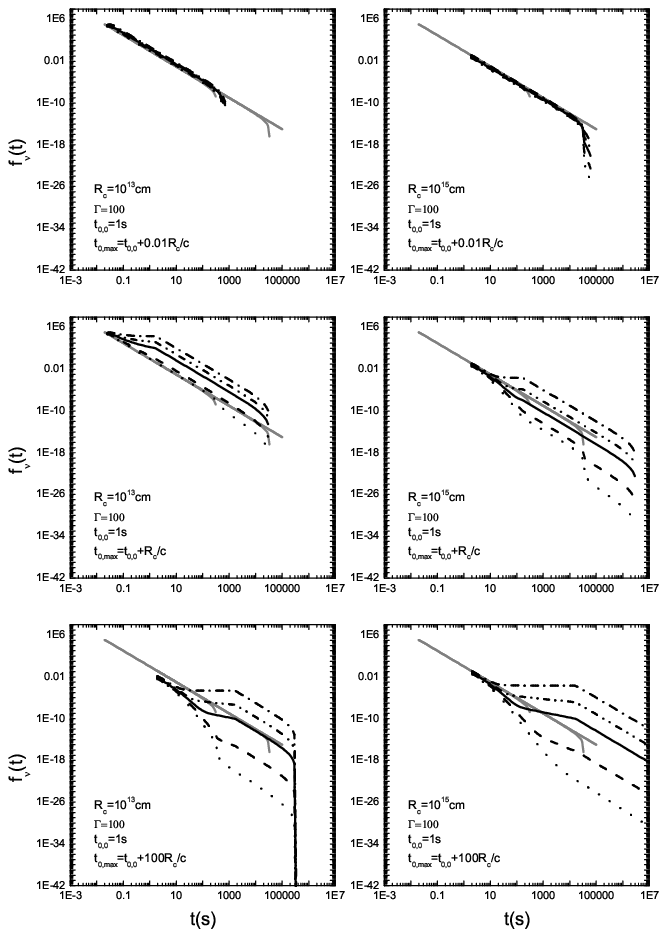}
\end{center}
\caption{Light curves of Fig. 4 replaced by those produced in the
case of $\Gamma =100$ and $t_{0,0}=1s$.} \label{Fig. 1}
\end{figure}

Upper panels of these figures show that when the power-law emission
is as short as $0.01$ times of the typical time scale of the
fireball radius (say, when $\Delta t_{0,0} = 0.01 R_{c}/c$) and the
Lorentz factor is not so large (say, not larger than 100), the light
curves are similar to those arising from the $\delta$ function
emission. This suggests that, in the framework of the curvature
effect, light curve characteristics of emissions with time scales as
short as $0.01$ times of $R_{c}/c$ and the Lorentz factor not larger
than 100 are hard to be distinguished from that of a $\delta$
function emission. (This is in agreement with what is shown in Fig.
2.)

Lower panels of these figures arise from longer duration of the
power-law emission ($\Delta t_{0,0} = 100R_{c}/c$). Some new
features appear. A remarkable one is the light curve with a
power-law decay curve followed by a shallow phase and then a steeper
power-law phase (type III). Connecting the two latter phases of this
kind of light curve is a remarkable time break (check the upper
three black lines of each lower panel of the figures). This tends to
happen when the intrinsic temporal power-law index is relatively
small. Otherwise, this kind of curve disappear (check the two lower
black lines in each lower panel of the figures).

When the duration of the power-law emission is not so large and not
so small (say, $\Delta t_{0,0} = R_{c}/c$), other forms of light
curves are observed (see the mid panels of these figures). In this
situation, when the Lorentz factor is large enough (say, $\Gamma
=100$), light curves of type III with shorter shallow phases appear
(see the mid panels of Figs. 6 and 7). This is expectable since in
the framework of the curvature effect the profile of a light curve
depends only on the ratio between the observational time scale and
the corresponding fireball radius time scale $R_{c}/c$ (see Qin et
al. 2004), and due to the contraction time effect (note that, the
face-on part of the fireball surface moves towards us when the
fireball expands) a certain observational time scale corresponds to
a longer co-moving time scale for a larger Lorentz factor.

\section{Example of application}

In our analysis above, we consider only a simple power-law emission,
for which the power-law index $\beta$ is assumed to be constant.
Expected from the model, the observed spectrum would be a constant
power-law with the same index and the rapid decay light curve would
be the well-known $t^{-(2+\beta)}$ curve. This constrains our
application, since the spectra of many X-ray afterglows of GRBs are
found to vary with time and the corresponding light curves are found
not to follow the $t^{-(2+\beta)}$ curve (see Zhang et al. 2007 and
the UNLV GRB Group web-site http://grb.physics.unlv.edu). Instead of
a power-law, many light curves are bent. To apply our model, one
must find a burst with its spectral index being constant and its
light curve following (or approximately following) the
$t^{-(2+\beta)}$ curve in its X-ray afterglow.

After checking the data provided in web-site
http://grb.physics.unlv.edu (up to March 25, 2008), we find that GRB
050219A might be one that fits our simple model. The data show, the
spectral index does not vary with time and its mean is $\beta =
0.907 \pm 0.051$. In addition, the first decay curve of the bust is
a power-law curve with its index being approximately $2+\beta$. This
phase is followed by a shallow one which starts at about $500s$.
Comparing this light curve with those presented in Fig. 3, we guess,
if it is due to the curvature effect, the fireball radius must be
larger than $R_{c} = 10^{13}cm$, otherwise the start time of the
shallow phase would be too small to meet the data (see the left
panels of Figs. 3a, 3b and 3c). If the radius is $R_{c} =
10^{15}cm$, then the Lorentz factor must be larger than $10$,
otherwise the start time of the shallow phase would be too large
(see the right upper panels of Figs. 3a, 3b and 3c). Revealed in
Fig. 3, there are four factors affecting the start time of the
shallow phase: the fireball radius $R_{c}$; the Lorentz factor
$\Gamma$; the intrinsic temporal power-law emission index
$\alpha_0$; and the start time of the intrinsic temporal power-law
emission $t_{0,0}$.

Available in the mentioned web-site, there are 75 data points in the
XRT light curve of GRB 050219A. As an example of fitting, we ignore
the three data points with the largest time scales since the gap
between them and the majority of the data set is too large and the
domain showing a constant spectral index does not cover them (see
http://grb.physics.unlv.edu) (in this way, one cannot tell if the
spectral index in the corresponding time scale is still constant).
With the rest 72 data points, we need only apply the equations
adopted in the discussion of the case of the temporal profile of the
intrinsic emission being a power-law function of time and the
power-law decay time being infinity. The equations adopted in
producing Fig. 3 are employed to fit the data set, where the term
$I_{p}\nu^{-\beta}$, which dominates the magnitude of the
theoretical curve, would be determined by fit.

Since both the fireball radius and the Lorentz factor are sensitive
to the time scale of the start time of the shallow phase, we deal
with them one by one. We first fix the Lorentz factor and assume it
to be $\Gamma= 100$, allowing $R_{c}$ and $\alpha_0$ to vary since
not only the start time of the shallow phase should be met but also
the power-law index of the shallow phase should be accounted for. In
addition, we take $t_{0,0}=1s$ since $t_{0,0}$ is less sensitive to
the start time of the shallow phase (see the right panels of Figs.
3a, 3b and 3c). The best fitting curve is shown in Fig. 8. One finds
that the XRT data of GRB 050219A can be roughly accounted for by a
power-law temporal emission from an expanding fireball surface. Note
that the corresponding fitting parameters are not important since
other possibilities exist (see the discussion below).

Next, we fix the fireball radius and take it as $R_{c} = 10^{15}cm$
(in fact, we take $R_{c}/v=(1/3)10^{5}s$ which corresponds to
$R_{c}\simeq 10^{15}cm$), allowing $\Gamma$ and $\alpha_0$ to vary.
Also, we take $t_{0,0}=1s$. The best fit is displayed in Fig. 9. It
shows that the result of the fit with a fixed fireball radius is
hard to be distinguished from that with a fixed Lorentz factor.
Therefore, the resulting fitting parameters are not important in
this stage of investigation.

There is a third choice: one could fix $\alpha_0$ and allow $R_{c}$
and $\Gamma$ to vary. We guess, it might yields a similar result.

\begin{figure}[tbp]
\begin{center}
\includegraphics[width=5in,angle=0]{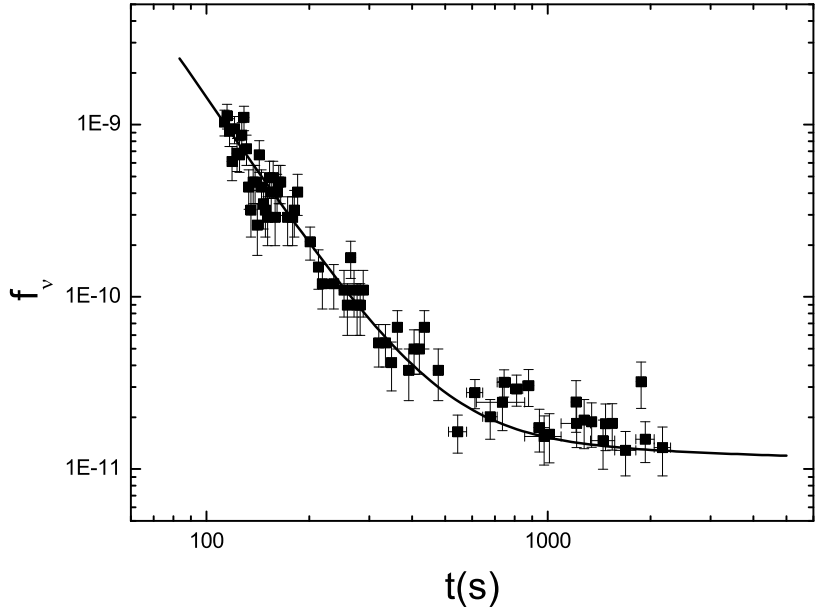}
\end{center}
\caption{XRT light curve of GRB 050219A. Equations (27) and
(32)-(35), which describe light curves arising from the intrinsic
emission with its temporal profile being a power-law function of
time and the power-law decay time being infinity, are employed to
fit the data, where we take $\Gamma= 100$. The solid line is the
best fit to the data. The corresponding fitting parameters are:
$R_{c}=1.05\times 10^{16} cm$, $\alpha_0=2.05$, and
$I_{p}\nu^{-\beta}=9.56\times 10^{-4}$ (see equation (27)). The
$\chi^2$ of the fit is $\chi^2_{dof}=1.39$.} \label{Fig. 1}
\end{figure}

\begin{figure}[tbp]
\begin{center}
\includegraphics[width=5in,angle=0]{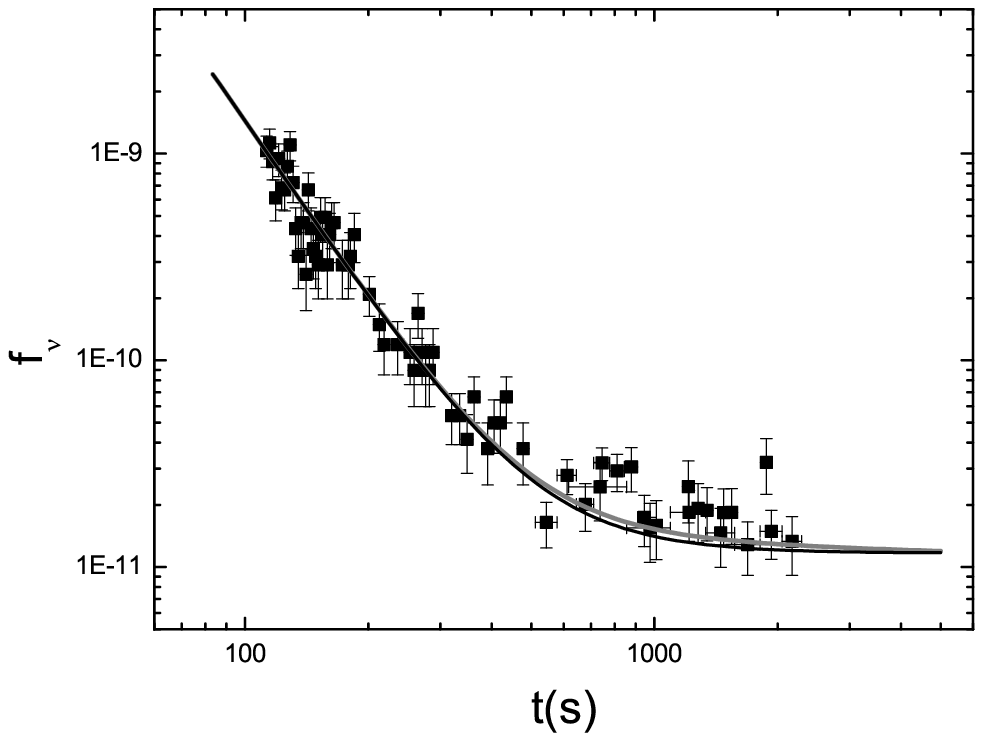}
\end{center}
\caption{Another fit to the XRT light curve of GRB 050219A, where we
take $R_{c}/v=(1/3)10^{5}s$. The equations adopted for the fit are
the same as those used in Fig. 8. The black solid line represents
the best fit and the grey solid line is the solid line in Fig. 8.
The corresponding fitting parameters are: $\Gamma=23.1$,
$\alpha_0=1.98$, and $I_{p}\nu^{-\beta}=8.88\times 10^{-4}$ (see
equation (27)). The $\chi^2$ of the fit is $\chi^2_{dof}=1.48$.}
\label{Fig. 1}
\end{figure}

\section{Discussion and conclusions}

We investigate in this paper how an intrinsic emission with a
power-law spectrum $I_{0,\nu }(t_{0},\nu _{0})=I_{0}(t_{0})\nu
_{0}^{-\beta }$ emitting from an expanding fireball surface gives
rise to an observed flux density when the full curvature effect is
considered. We find that, if the power-law spectrum of the intrinsic
radiation holds within the energy range that corresponds to the
observed energy channel due to the Doppler shifting, the resulting
spectrum would be a power-law as well and the index will be exactly
the same as that in the intrinsic spectrum, regardless the real form
of the temporal profile of the intrinsic emission. Accompanied with
the power law spectrum of index $\beta$ is a power law light curve
with index $2+\beta $, expected by the curvature effect, which was
known previously (see Fenimore et al. 1996; Kumar \& Panaitescu
2000). This light curve could be observed if the intrinsic emission
is extremely short or if the emission arises from an exponential
cooling.

In particular, we assume and consider a power-law cooling emission
in the co-moving frame (for this emission, the intrinsic temporal
profile is a power-law). We find that, if the power-law decay time
being infinity, due to the contribution of the power-law cooling in
the co-moving frame, the observed light curve influenced by the full
curvature effect contains two phases: one is a rapid decay phase
where the light curve well follows the well-known $t^{-(2+\beta)}$
curve, and the other is a shallow decay phase where the light curve
is obviously shallower than that in the rapid decay phase. If the
power-law decay time is limited, there would be several kinds of
light curve. A remarkable one among them contains three power-law
phases (see Figs. 4-7): the first is a rapid one with its index
being equal to or larger than that of the $t^{-(2+\beta)}$ curve;
the second is a shallow decay one with its index being obviously
smaller than that in the first phase; and the third is a rapid decay
one with its index being equal to or less than that of the first
phase. It might be possible that, some of the GRBs containing such
features in their afterglow light curves are due to expanding
fireballs or face-on uniform jets (see e.g. Qin et al. 2004)
emitting with a power-law spectrum and a power-law cooling (being
infinity or limited). In the view of co-moving observers, the
dynamic process of the merger of shells would be somewhat similar to
that occurred in the external shocks (the main difference is that in
the case of inner shocks, a co-moving observer observes only a
limited volume of medium for which the density would evolve with
time due to the enhancement of the fireball surface). Based on this
argument, we suspect that the intrinsic emission of some of those
bursts possessing in their early X-ray afterglows a rapid decay
phase soon followed by a shallow decay phase and then a rapid decay
one might be somewhat similar to the well-known standard forward
shock model (Sari et al. 1998; Granot et al. 1999); while for some
of the bursts with a rapid phase followed by a shallow phase in
their late X-ray afterglows the emission might be that of the
standard forward shock model influenced by the curvature effect.
Necessary conditions for perceiving this mechanism include: a)
during the period concerned, the spectral index should be constant;
b) the temporal index in the first phase should be equal to or
larger than that of the $t^{-(2+\beta)}$ curve.

As an example of application, we employ the XRT data of GRB 050219A
to perform a fit since the spectral index $\beta$ of this burst does
not vary with time and the first decay phase of its light curve is a
power-law one with its index being approximately $2+\beta$. The
result shows that the XRT data of this burst can be roughly
accounted for by a power-law temporal emission from an expanding
fireball surface. Since there exist various possibilities,
parameters obtained by the fit are not unique. To determine the
parameters, we need other independent estimations. According to the
analysis above, a reliable value of the fireball radius would be
obtained if one observes a ``cut-off'' feature following the
$t^{-(2+\beta)}$ curve in the case of a constant spectral index
$\beta$. Nevertheless, the start time of the shallow phase could
raise a limit to the fireball radius (see Figs. 3-7). For GRB
050219A, if its XRT data are indeed due to the curvature effect, its
radius corresponding to this emission must be larger than $R_{c} =
10^{13}cm$. We have checked that taking $R_{c}/v=(1/3)10^{4}s$
(which corresponds to $R_{c}\simeq 10^{14}cm$), $\Gamma=8$ and
$\alpha_0 = 2.157$ can also roughly account for the data. As the
Lorentz factor is so small in this situation, we conclude that, if
the X-ray afterglow of GRB 050219A does arise from the emission of
an expanding fireball surface, the radius of the fireball associated
with this emission would not be much less than $R_{c}= 10^{14}cm$,
otherwise $\Gamma$ would be too small to be regarded as a
relativistic motion.

Why a shallow phase emerges due to the curvature effect? We guess,
while the first phase is dominated by the geometric effect and
therefore obeys the $t^{-(2+\beta)}$ curve, in the shallow phase the
intrinsic emission overcomes the geometric effect and dominates the
light curve observed. One might notice the $\alpha _{0}=2$ lines
(the dash dot lines) in Fig. 3 --- the shallow phase curve of these
lines is parallel to the time axis. As the radius grows linearly
with time when a constant Lorentz factor is assumed (see Qin 2002),
the emission from the fireball surface of a certain solid angle
increases as a square of time (the area of the surface is
proportional to $R^2$). This in turn makes the total emission of
$I_{0} \propto t_{0}^{-2}$ from the surface becoming constant. When
the intrinsic emission overcomes the geometric effect in the shallow
phase, one cannot expect a light curve of $t^{-\alpha_0}$, but
instead, we expect that of $t^{-(\alpha_{0}-2)}$ (see Fig. 3).

It is known that a $\delta $-function intensity approximates the
process of an extremely short emission. This will occur when the
corresponding fireball shells are very thin and the cooling time is
relatively short compared with the curvature time scale (for the
time scale of the curvature effect, see Kocevski et al. 2003 and Qin
\& Lu 2005). Two light curve characteristics are associated with a
quasi-$\delta$-function emission. The first is a strict power-law
decay curve with index $2+\beta$. The second is the limited time
range of this curve. If the cooling time is not so short but it is
an exponential one, then these characteristics are also expected
(see Fig. 2). Note that the exponential cooling time does not last
the $t^{-(2+\beta)}$ curve to a much larger time scale when the
cooling itself is not very large (say, in the case of $R_{c} =
10^{15}cm$, $\sigma _{d}<100s$; see Fig. 2). Thus, one can estimate
the fireball radius from bursts possessing these characteristics
(note that, the time scale of the $t^{-(2+\beta)}$ curve is
independent of the Lorentz factor; see equations (16) and (18)). For
candidates of this kind of burst, we propose to fit the spectrum
with $\nu ^{-\beta }$ and the light curve with
$[T_{D}-(t-T_{0})](t-T_{0})^{-(2+\beta )}$, where both $T_{D}$ and
$T_{0}$ are free parameters. When the fitting is good enough, we say
that the intrinsic fireball emission is likely very short or the
cooling is an exponential one and the corresponding fireball radius
is $R_{c}\simeq vT_{D}$. When the expansion of the fireball is
relativistic, we get $R_{c}\simeq cT_{D}$. Therefore, via this
method, one obtains at least the upper limit of the fireball radius
as long as the intrinsic emission is extremely short, or the cooling
is an exponential one, and the intrinsic spectrum is a power-law.

In the case of the intrinsic temporal power-law emission, when its
temporal index is large enough ($\alpha _{0}>2+\beta$), there would
be a ``cutoff'' curve located exactly at the same time position of
the speedily falling off tail in the light curve of a
$\delta$-function emission. This feature could be used to estimate
the fireball radius as well. Presented in Zhang et al. (2007),
several bursts seem to possess this ``cutoff'' feature: GRB050724,
GRB060211A, GRB060218, GRB060427, GRB060614, GRB060729 and
GRB060814. If the proposed interpretation can be applied, their
radius would be that ranging from $10^{13}cm$ to $10^{15}cm$. At
lease one reason prevents us to reach such a conclusion. The spectra
of these bursts happen to vary quite significantly within the light
curves associated with this feature. This conflicts with what we
assume in this paper (we assume a constant intrinsic spectrum). We
thus appeal further investigation of this issue taking into account
the variation of the intrinsic spectrum, which might tell us whether
the ``cutoff'' feature remains and/or its properties are maintained.

Displayed in literature, many Swift bursts are found to possess a
bent light curve instead of a strict power-law one, in the early
X-ray afterglow (see, e.g., Chincarini et al. 2005; Liang et al.
2006; Nousek et al. 2006; O'Brien et al. 2006). In our analysis
above, we seldom get bent light curves. This must be due to the fact
that the model concerned is too simple, where we consider only
emissions with constant spectra. When the intrinsic spectrum varies
with time, one would expect bursts with both variable spectra and
bent light curves (the well-known $t^{-(2+\beta)}$ curve suggests
that light curves of fireballs are strongly affected by the
corresponding emission spectra). Indeed, we find that both variable
spectra and bent light curves happen to appear in the same period
for many Swift bursts (see Zhang et al. 2007 and the UNLV GRB Group
web-site http://grb.physics.unlv.edu).

Since for some bursts their early X-ray afterglow spectra evolve
with time while for some others their spectra have no significant
temporal evolution (Zhang et al. 2007; Butler \& Kocevski 2007b), we
suspect that there might be two kinds of mechanism accounting for
the X-ray afterglow emission. It seems likely that the observed
variation of spectra is due to an intrinsic spectral evolution. The
intrinsic spectral evolution would probably lead to deviations of
the light curves studied above (those studied in Figs. 3-7). We thus
suspect that the bursts with no spectral evolution might have
``normal'' temporal profiles, while others might exhibit somewhat
``abnormal'' profiles. This seems to be the case according to Figs.
1-3 in Zhang et al. (2007) and Figs. 7-8 in Butler \& Kocevski
(2007b).

Our simple model tends to account for the kind of bursts that their
X-ray afterglow spectra do not evolved with time. However, for many
bursts with roughly constant spectra and power-law light curves, the
curves are too shallow to be accounted for by the $t^{-(2+\beta)}$
curve (see http://grb.physics.unlv.edu). Our model seems too simple
to account for the majority of XRT light curve data of Swift bursts.
It is therefore necessary to explore more complicated cases. For
example, a variant Lorentz factor (which is expectable when the
intrinsic emission is long enough) might play a role. Would it
affect the slope of the decaying curve? We are looking forward to
see more investigations on this issue in the near future.

Before ending this paper, we would like to point out that quantity
$t_{0,c}$ is the co-moving time measured by a co-moving observer
when the fireball radius reaches $R_c$ (see Qin 2002). Note that,
$t_{0,0} > t_{0,c}$. Therefore, when assigning $t_{0,c} = 0$,
$t_{0,0} = 1s$ means $1s$ co-moving time has passed after $R = R_c$.
When one analyzes the emission associated with $R_c = 10^{15} cm$,
$t_{0,0} = 0$ refers only the emission at $t_{0,c} = 0$ which is the
co-moving time when $R = R_c$. Although we take quite small values
of $t_{0,0}$ in the above analysis, it does not correspond to early
emission when we adopt $R_c = 10^{15} cm$ or $R_c = 10^{13} cm$.
Therefore, our analysis on emission from fireballs with $R_c =
10^{15} cm$ does not put forward any constraint on the prompt
emission. The conclusion that characteristics of the prompt emission
of bursts with shallow decay phase are similar to those without
shallow decay phase obtained recently by Liang et al. (2007) are not
violated by our findings.

\acknowledgments

This work was supported by the National Science Fund for
Distinguished Young Scholars (10125313), the National Natural
Science Foundation of China (No. 10573005), and the Fund for Top
Scholars of Guangdong Provience (Q02114). We also thank the
financial support from the Guangzhou Education Bureau and Guangzhou
Science and Technology Bureau.

\end{document}